\documentstyle[psfig,draft]{mn}

\def\gs{\mathrel{\raise0.35ex\hbox{$\scriptstyle >$}\kern-0.6em
\lower0.40ex\hbox{{$\scriptstyle \sim$}}}}
\def\ls{\mathrel{\raise0.35ex\hbox{$\scriptstyle <$}\kern-0.6em
\lower0.40ex\hbox{{$\scriptstyle \sim$}}}}
\def\ls{\mathrel{\hbox{\rlap{\hbox{\lower4pt\hbox{$\sim$}}}\hbox{$<$}}}}
\def\gs{\mathrel{\hbox{\rlap{\hbox{\lower4pt\hbox{$\sim$}}}\hbox{$>$}}}}
\def\ergs {{\rm erg} \, {\rm s}^{-1}}
\def\mnras {{\sc MNRAS}}
\def\apj {ApJ}

\def\apjl {ApJL}
\def\aj {AJ}
\def\aaps {A\&AS}

\title[LARCS I: Photometric Catalogues]
      {The Las Campanas/AAT Rich Cluster Survey I: Precision and Reliability of the Photometric Catalogue}

\author[K.\,A.\ Pimbblet et al.]
       {Kevin A.\ Pimbblet,$^{\! 1}$ 
        Ian Smail,$^{\! 1}$ Alastair C.\ Edge,$^{\! 1}$ 
        Warrick J.\ Couch,$^{\! 2}$ 
  	\and Eileen O'Hely,$^{\! 2}$ 
	and Ann I.\ Zabludoff.$^{3}$
        \vspace*{1mm}\\
        $^1$ Department of Physics, University of Durham, South Road,
        Durham, DH1 3LE, UK\\
	$^2$ School of Physics, University of New South Wales, Sydney, 
	NSW 2052, Australia\\
	$^3$ Steward Observatory, University of Arizona, Tucson, 85721, USA}

\pagerange{000--000}

\begin{document}

\maketitle

\begin{abstract}
The Las Campanas Observatory and Anglo--Australian Telescope Rich
Cluster Survey (LARCS) is a panoramic imaging and spectroscopic survey
of an X-ray luminosity-selected sample of 21 clusters of galaxies at
$0.07<z<0.16$.  CCD imaging was obtained in $B$ and $R$ of typically
2-degree wide regions centred on the 21 clusters, and the galaxy sample
selected from the imaging is being used for an on-going spectroscopic
survey of the clusters with the 2dF spectrograph on the 
Anglo-Australian Telescope.
This paper presents the reduction of the imaging data and the 
photometric analysis used in the survey.  
Based on an overlapping area of $12.3$ square degrees, we compare
the CCD-based LARCS catalogue with the
photographic-based galaxy catalogue used for the input to the
2dF Galaxy Redshift Survey (2dFGRS) from the Automated Plate 
Measuring Machine(APM)  
to the completeness of the GRS/APM catalogue, $b_J=19.45$.  
This comparison confirms the reliability of the photometry across
our mosaics and between the clusters in our survey.  
This comparison also provides useful information about
the properties of the GRS/APM.  
The stellar contamination in the GRS/APM
galaxy catalogue is confirmed to be around 5--10 percent, as 
originally estimated.  However, using the superior sensitivity 
and spatial resolution in the LARCS survey we find
evidence for four distinct populations of galaxies
that are systematically omitted from the GRS/APM catalogue.  
The characteristics of the
`missing' galaxy populations are described, reasons for their absence
examined and the impact they will have on the
conclusions drawn from the 2dF Galaxy Redshift Survey are discussed.
\end{abstract}

\begin{keywords}
surveys, catalogues, galaxies: photometry
\end{keywords}

\section{Introduction}
Panoramic surveys based upon systematic photographic imaging of
large areas of the sky have under-pinned a large fraction of astrophysical
research since the 1950s, with the three main Schmidt telescopes,
at Palomar, Siding Springs and La Silla, completing at least 14 major
surveys over this period.  The full exploitation of these observations
has been achieved through the digitization of the photographic plates
on facilities such as the Automated Plate Measuring Machine 
(APM, Kibblewhite et al.\ 1984) at Cambridge, or COSMOS at 
Edinburgh (MacGillivray \& Stobie, 1984).

While they cover impressively large areas, the photographic surveys
have several drawbacks which makes them unsuitable for some applications.
In particular, the low sensitivity and non-linear response
of standard photographic plates means that the surveys have relatively
bright surface brightness limits and require significant effort to
reliably calibrate the magnitude scale over the whole range detected.
Indeed repeated scans of individual plates using the measuring
machines suggests a scatter in the measured magnitudes for sources
of at least $\sim 0.04$\,mags (Maddox et al.\ 1990a), 
before other contributions are included.
The relatively poor spatial resolution achieved on the Schmidt plates also
limits the reliability of star--galaxy separation, placing additional
restrictions on the questions which can be addressed with these data.

Panoramic imaging surveys using CCDs circumvent many of the drawbacks
of photographic surveys by virtue of their high quantum efficiency
and good linearity of CCD devices.  Large format CCDs (and mosaic cameras) 
are therefore now being used to undertake wide-field surveys, whose
depth, resolution and photometric precision far exceeds those 
achieved using photographic plates.

This paper describes one such project -- the Las Campanas/AAT Rich Cluster
Survey (LARCS).  LARCS is a long-term project to study a 
statistically-reliable sample of the most luminous X-ray clusters 
at intermediate
redshifts ($z=0.07$--0.16) in the southern hemisphere.  The goals
of the project are to understand the influence of environment on the
characteristics of galaxies, such as luminosity, star formation history
and morphology.  To achieve this 
we map the photometric, spectroscopic and
dynamical properties of galaxies in rich cluster environments at $z\sim
0.1$, tracing the variation  in these properties from the high-density
cluster cores out into the surrounding low-density field, beyond the
turn-around radius ($\gs 10$\,Mpc)\footnote{Throughout this 
work values of $H_0 = 50 h$  
km s$^{-1}$ Mpc$^{-1}$ and $q_o=0.5$ have been adopted.}.
For the most massive clusters at
$z\sim 0.1$, the turn-around radius corresponds to roughly 1\,degree and
therefore  we require panoramic CCD imaging covering 2-degree diameter
fields, as well as spectroscopic coverage of similar fields.  The former
is achieved by mosaicing CCD images
from the 1-m Swope telescope at Las Campanas Observatory, 
while the latter comes from the
subsequent spectroscopic follow-up with the 400-fibre 2dF multi-object
spectrograph on the 3.9-m Anglo-Australian Telescope (AAT).

To ensure that the results from the survey can be 
reliably compared to theoretical models,
well-defined selection criteria are used to identify the cluster sample
used. 
The LARCS sample is selected from
the X-ray brightest Abell Clusters (XBACs) 
catalogue of Ebeling et al.\ (1996) constructed from the
{\it ROSAT} All-Sky Survey.  XBACs is an all-sky, X-ray flux limited,
sample of 242 Abell clusters and is effectively
a complete sample of richest clusters at $z \ls 0.16$ (Ebeling et al. 1996).
We therefore
restrict our sample to the most X-ray luminous ($L_{X} \geq 3.7 \times
10^{44}$\,erg\,s$^{-1}$) southern clusters (with $\delta \leq 10 \deg$,
so accessible from the AAT) lying in the redshift range $0.07 \leq z
\leq 0.16$.  These selection criteria result in 53 clusters in the XBACs
sample and from these a random subsample of 21 is selected for our survey.
LARCS has a sufficiently large sample that it should provide a 
statistically-reliable view of the properties of both the cluster 
galaxies and how
these relate to the characteristics of the clusters themselves.
A summary of the 21 galaxy clusters is presented in 
Table \ref{tab:clusters}, together with the current state of observations
made.

%
%
\begin{table*}
\begin{center}
\caption{The 21 clusters in the LARCS sample
and the multiwavelength observations so far obtained.  
The column headed `Passbands'
indicates the broad band filters used to observe each cluster. 
Other observations indicate {\it ROSAT} imaging (with the {\it PSPC}
or {\it HRI} instruments), {\it XMM-Newton EPIC}
time that has been allocated and observations undertaken
with the 2dF spectrograph at the AAT.\hfil}
\begin{tabular}{lcccclcl}
\noalign{\medskip}
\hline
Cluster & R.A.\ & Dec.\ & $z$ & $L_X$ & Passbands & Diameter  & Other Observations \\
         &  \multispan2{\hfil(J2000)\hfil} & & ($\ergs$) & & (deg.) & \\
\hline
A22   & 00 20 38.64 & $-$25 43 19 & 0.131 &  5.31 & $B/R$ & 2.0 & 2dF \\
A550  & 05 52 51.84 & $-$21 03 54 & 0.125 &  7.06 & $B/R$ &  2.0 & \\
A644  & 08 17 25.20 & $-$07 31 41 & 0.071 &  7.92 & $B/R$ &  2.0 &{\it PSPC, HRI, EPIC} \\
A1084 & 10 44 30.72 & $-$07 05 02 & 0.134 &  7.42 & $B/R$ &  2.0 &{\it EPIC} \\
A1285 & 11 30 20.64 & $-$14 34 30 & 0.106 &  5.47 & $B/R$ &  2.0 & \\
A1437 & 12 00 25.44 & +03 21 04 & 0.133 &  7.72 & $B/R$ & 2.0 & {\it HRI} \\
A1650 & 12 58 41.76 & $-$01 45 22 & 0.084 &  7.81 & $B/R$ & 2.0 & {\it HRI} \\
A1651 & 12 59 24.00 & $-$04 11 20 & 0.084 &  8.25 & $B/R$ & 2.0 & {\it PSPC} \\
A1664 & 13 03 44.16 & $-$24 15 22 & 0.127 &  5.36 & $B/R$ & 2.0 & {\it PSPC, HRI} \\
A2055 & 15 18 41.28 & +06 12 40 & 0.102 &  4.78 & $B/R$ &  2.0 &{\it HRI, EPIC}, 2dF \\
A2104 & 15 40 06.48 & $-$03 18 22 & 0.155 &  7.89 & $B/R$ & 2.0 & {\it PSPC, EPIC}, 2dF \\
A2204 & 16 32 46.80 & +05 34 26 & 0.152 & 20.58 & $B/R$ & 2.0 & {\it PSPC}, 2dF \\
A2597 & 23 25 16.56 & $-$12 07 26 & 0.085 &  7.97 & $B/R$ & 2.0 & {\it PSPC, HRI} \\
A3112 & 03 17 56.40 & $-$44 14 17 & 0.070 &  7.70 & $B/R$ & 2.0 & {\it PSPC} \\
A3378 & 06 05 52.80 & $-$35 18 04 & 0.141 &  6.87 & $B/R$ & 2.0 & {\it HRI} \\
A3888 & 22 34 32.88 & $-$37 43 59 & 0.151 & 14.52 & $B/R$ & 2.0 & {\it PSPC, EPIC}, 2dF \\
A3921 & 22 49 59.76 & $-$64 25 52 & 0.095 &  5.40 & $B/R$ & 2.0 & {\it PSPC}, 2dF \\
\noalign{\smallskip}
A2811 & 00 42 07.92 & $-$28 32 10 & 0.108 &  5.43 & $U/B/R/K$ & 0.6 & {\it HRI} \\
A2345 & 21 26 58.56 & $-$12 08 28 & 0.176 &  9.93 & $U/B/R/K$ & 0.6 & {\it HRI, EPIC} \\
A3814 & 21 49 06.48 & $-$30 41 53 & 0.117 &  3.85 & $U/B/R/K$ & 0.6 & {\it HRI} \\
A2496 & 22 50 55.92 & $-$16 23 35 & 0.122 &  3.71 & $U/B/R/K$ & 0.6 & {\it HRI}		 \\
\hline
\noalign{\smallskip}
\end{tabular}
  \label{tab:clusters}
\end{center}
\end{table*}

The project has completed panoramic ($2\times 2$ degree$^2$) $B$ and $R$
broad-band imaging  of the cluster sample from the 1-m Swope telescope at Las Campanas
Observatory, Chile, totaling over 70 square degrees of
sky (O'Hely et al.\ 1998, O'Hely 2000).  
These images are used to select galaxies
for the on-going spectroscopic survey using 2dF. The final goal is to
obtain spectra for roughly 20,000 galaxies in the 21 clusters providing
an unprecedented view of the dynamics of rich clusters and their galaxy
populations within a region encompassing the core, halo and in-fall
regions of the clusters.

This paper presents the details of the photometric reduction, calibration
and analysis of the CCD imaging used in LARCS.  In addition, as a test
of the LARCS catalogues we compare our catalogues in the fields of four
clusters that overlap with the plate-based galaxy catalogue from the
APM, (Maddox et al.\ 1990a, 1990b) 
which is the basis of the 2dF Galaxy Redshift Survey 
(2dFGRS, Colless et al.\ 1998; Maddox et al.\ 1998).

The plan of the paper is as follows: In \S2 we discuss the imaging
observations obtained for our survey, data reduction methods and quality
tests.  We test both the photometric properties and the star-galaxy
separation of our catalogues.  In \S3 we compare the galaxy catalogues
derived from our CCD imaging with those from the photographic APM
catalogue used by the 2dFGRS (referred to as the GRS/APM 
catalogue in the following).  
\S4 deals with
the properties of the galaxy populations which are missing from the
GRS/APM catalogue -- including both compact, high surface brightness
galaxies and lower surface-brightness galaxies, examining their
magnitude, spatial and colour distributions. 
We summarize our main conclusions in \S5.

\section{Imaging Observations, Reduction and Analysis}

Analysis of the clusters is based upon panoramic CCD imaging from
the 1-m Swope telescope.
High quality broad-band $B$ and $R$ CCD images of all the cluster were
obtained over the course of three years, 1996--1998.  
These observations are summized in Table~\ref{tab:clusters}.
The observations
employed a thinned $2048 \times 2048$, $24 \mu$m pixel Tektronics 
CCD giving a pixel scale of 0.696$''$/pixel and a field of 
view for each exposure of $23.76' \times 23.76'$.  
With 44$''$ overlaps the total coverage of the $5\times
5$ mosaic of pointings is $\sim 2$\,degrees diameter (the corners of
the mosaic are omitted to give a total of 21 pointings per cluster).
The total exposure time per pointing is 500\,s in $B$ and 400\,s 
in $R$, each split into two spatially-offset sub-exposures
to facilitate cosmic-ray rejection and the removal of a small number of
cosmetic features.

Here we have chosen to discuss in detail our 
optical analysis of
four of the clusters: Abell 22, 1084, 1650 and 1651, which are also
included within the region surveyed by the 2dFGRS.  
These clusters were observed in two runs:
March 16$^{\rm th}$ to 22$^{\rm nd}$ 1996 and August 19$^{\rm th}$
to 21$^{\rm st}$ 1996.  The details of these observations are in
Table~\ref{tab:obs}.

%
%
\begin{table*}
\begin{center}
\caption{Log of LCO imaging observations made for the four
clusters used in this work. \hfil}
\begin{tabular}{lccccl}
\noalign{\medskip}
\hline
Cluster & Date & Passband & N(point) & Seeing & Photometric? \\
     &    (1996) &      &          & ($''$)   & \\
\hline
Abell~22   & Aug 20 & $R$ & ~3 & 1.0--1.1 & yes \\
           & Aug 21 & $R$ & ~6 & 1.1--1.3 & yes \\
           & Aug 22 & $R$ & 12 & 1.0--1.4 & yes \\
           & Aug 21 & $B$ & 15 & 1.0--1.1 & yes \\
           & Aug 22 & $B$ & ~6 & 1.0--1.1 & yes \\
Abell~1084 & Mar 20 & $R$ & 10 & 1.1--1.8 & no \\
           & Mar 21 & $R$ & 11 & 1.0--1.2 & partial \\
           & Mar 17 & $B$ & 10 & 1.4--1.9 & yes \\
           & Mar 18 & $B$ & 11 & 1.1--2.1 & yes \\
Abell~1650 & Mar 20 & $R$ & 21 & 1.0--1.2 & partial \\
           & Mar 16 & $B$ & 15 & 1.1--1.7 & yes \\
           & Mar 22 & $B$ & ~6 & 1.2--2.3 & yes \\ 
Abell~1651 & Mar 21 & $R$ & 21 & 1.0--1.1 & partial \\
           & Mar 17 & $B$ & 21 & 1.1--1.4 & yes \\
\noalign{\smallskip}  \hline
\end{tabular}
  \label{tab:obs}
\end{center}
\end{table*}

The imaging data are reduced in a standard manner using packages within
{\sc iraf}, steps undertaken include debiasing, preliminary flatfielding
with twilight flats, creation of super-flats from independent stacks of
the science frames and the application of these to the science frames,
alignment and co-addition of data using a cosmic-ray reject algorithm.
The accuracy of the alignment is better than one pixel in all cases.  

To test the precision of the flatfielding, the {\sc iraf} task
{\sc imstat} is used to obtain the mean background levels at 
various locations on each mosaic tile.  
The variation in the sky background is found to be $\leq 1$ percent 
in all cases.

Before constructing the object catalogues it is necessary to match the seeing
of the $B$ and $R$-band tiles so that reliable
aperture colours could be measured. 
The matching is achieved by degrading the tile with the
better seeing to that of the worse.
A first pass of SExtractor (Bertin \& Arnouts 1996) is used
to obtain a rough list of stellar sources on the $R$-band tiles,
by taking objects with {\sc class\_star} $ > 0.9$ (see \S2.3).
The seeing of the $B$ and $R$ tiles is estimated
from the median value of the FWHM of these sources measured with
the {\sc imexamine} task.  
An initial estimate of $\sigma$, the value used to
convolve the better seeing tile (typically the
$R$-band images) to the worse one, is made and
applied using the {\sc gauss}
task to create a new, degraded tile.  The median seeing for
the stars on the degraded tile is then measured using {\sc imexamine}.
The process is repeated iteratively to determine the
correct value for $\sigma$ (within the
1-$\sigma$ scatter of the median FWHM) to match the seeing on the two tiles. 
Before convolution, differences of up to $1''$ may be present between the
median FWHM of the $R$- and $B$-band tiles.  
After convolution the difference between the 
median seeing is $\ls 0.1''$ and colours can be reliably measured.

SExtractor is used to automatically
analyse the tiles, detect sources and parametrize them.  We catalogue
the $R$-band frames, detecting sources having more than 12 contiguous
pixels each $3\sigma$ of the sky above the background, 
equivalent to $\mu_R\sim23.3$\,
mag.\ arcsec$^{-2}$.  The resulting catalogues have $5\sigma$ detection
limits for galaxies of $R=21.1$--21.3. Spurious sources,
such as satellite trails and diffraction spikes from bright stars,
are cleaned manually from the resultant catalogues by visual inspection.

To derive colours for these sources {\sc phot} within {\sc iraf} is used
and colours are measured in $4''$ diameter apertures (typically $\sim10$\,kpc
at the cluster redshifts) from the seeing-matched frames. 
These colours are used in the following analysis to
convert the $R$-selected sample to $B$-band to compare to
the GRS/APM.  

\subsection{Photometry}

%
%
%
\begin{figure*}
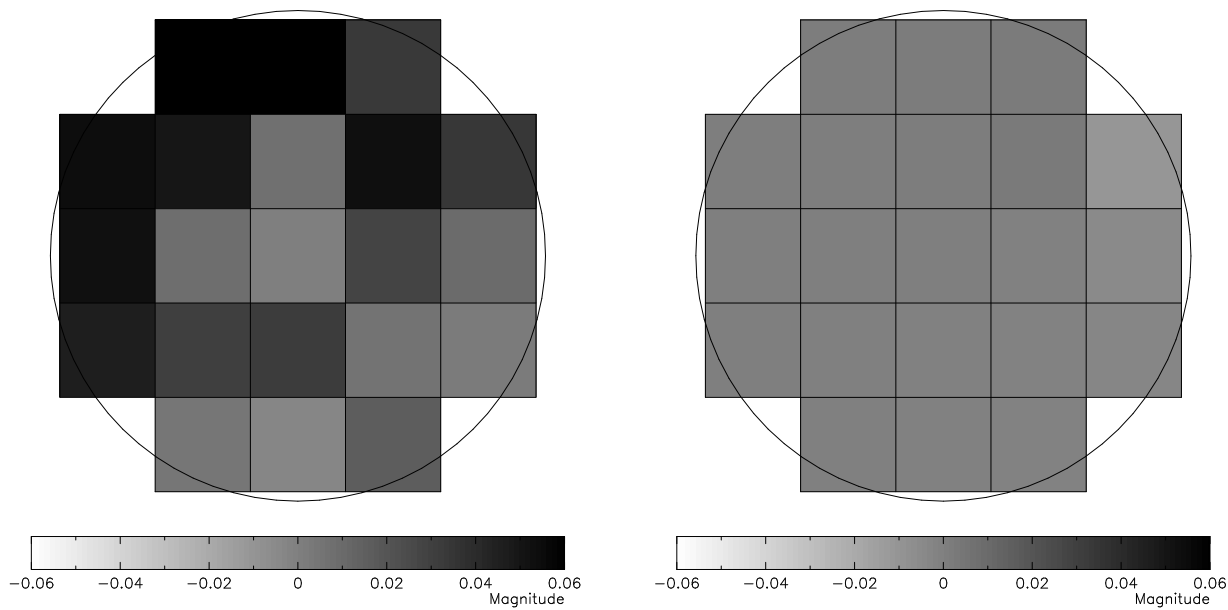

\centerline{
\psfig{file=BEFORE.ps,angle=0,width=3.0in} 
\hspace*{0.3in}
\psfig{file=AFTER.ps,angle=0,width=3.0in}
}
\caption{Left-- Grey-scale of the zeropoint offsets
between the different tiles within the $R$-band mosaic of Abell~1651.
The range of offsets is given by the scale at the bottom.  The mosaic
is non-photometric and shows a range of over 0.06 magnitudes
across the mosaic and an rms of 0.02. 
The circle has a diameter representing 2 degrees on the sky.
Right-- The same field after application
of the iterative corrections from Glazebrook (1994) --
the tile-to-tile rms is now reduced to $\leq 0.005$\,mag.}
  \label{fig:diffs}
\end{figure*}

The science frames are calibrated from the observations
of standard stars selected from Landolt (1992), interspersed 
throughout each night.  
Typically we observed over 100 standard stars each night
in each broad-band filter throughout the observing runs.
We fit for colour terms each night from 
observations of the standards taken at the same airmass.
Extinction is accounted for by fitting to the variation
with airmass of the standard stars photometry in one night.
On photometric nights, the variation in extinction coefficient
and colour term are all within $1\sigma$ of each other.
The photometric solutions for the photometric nights 
(Table~\ref{tab:obs}) are typically accurate to better 
than 0.03\,mags.  

As some of the observations are made in non-photometric conditions it is
necessary to independently calibrate such pointings.
To achieve this, the photometry of sources appearing in the
44$''$ inter-tile overlap regions is used to determine the magnitude offset
between the non-photometric and photometric tiles in each passband.
Only those sources which are unsaturated, not heavily blended or
close to the chip's edge
are used to obtain these values.  To optimally employ magnitude offset
information across the whole mosaic we use the method of Glazebrook (1994,
G94).  Using a single step algorithm, G94 generates appropriate magnitude
offsets to be applied to the non-photometric tiles, whilst keeping
the zero-points of the photometric tiles fixed.  After application of
G94 we estimate that the typical tile to tile variation in zero-points
is $\leq 0.006$\,mags, as estimated from the scatter in the magnitude
offsets from the duplicated sources in the overlapping regions.  The
maximum deviations between tiles is $\sim 0.015$\,mags.
The photometric zeropoint errors of 0.03\,mags are the 
dominant ones and the tile-to-tile variation in zero-points is small by 
comparison.

Figure~\ref{fig:diffs} shows the zeropoint offsets between each
pointing of an example (non-photometric) mosaic as a grey-scale
before and after correction using the G94
algorithm.  After calibration the entire mosaic 
is close to uniform, with a tile-to-tile rms of only 
0.005\,mags.

Once both passbands are calibrated, the final catalogue is
constructed by combining the catalogues for the individual 21 
tiles across the entire mosaic.  Duplicate sources (from the
overlapping regions) are removed by averaging their critical
SExtractor parameters (e.g.\ total magnitude).

\subsection{Astrometry}

Astrometry is performed individually on each pointing of the mosaic.
Approximately 60 bright ($B<18.0$) stars per pointing are tied to the
positions from the APM (http://www.ast.cam.ac.uk/apm/).  
The positions of the rest of the sources (galaxies and uncatalogued stars)
are then obtained using the {\sc astrom} package.
As a quality test, we compare the astrometrical solution of sources
within the 44$''$ overlap regions of the 2 degree mosaics.
Our internal accuracy is found to be better than $\leq 0.3''$,
adequate for the needs of spectroscopy on 2dF.  

\subsection{Star--Galaxy Separation}

Accurate star--galaxy separation is essential in extragalactic surveys.  
Robust separation is vital for minimizing stellar contamination,
while at the same time retaining compact galaxies and optimising the
efficiency of spectroscopic follow-up of the survey.

\begin{figure}
\centerline{\psfig{file=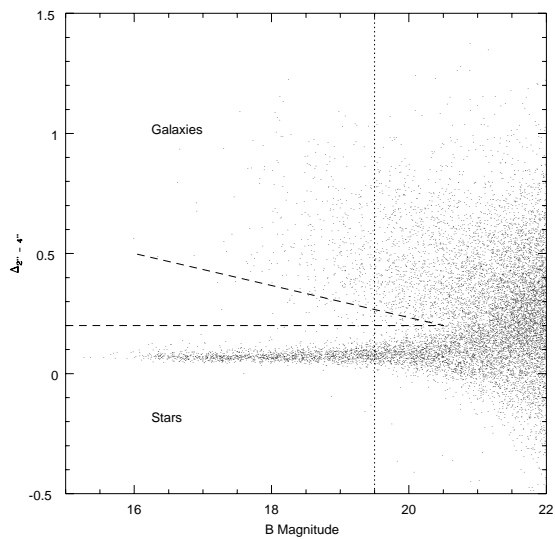,angle=0,width=3.0in}}
  \caption{
Difference between the 2.0$''$ and 4.0$''$ aperture magnitudes
($\Delta_{2''-4''}$) versus total magnitudes for all 31,851 sources
within the LARCS Abell~22 catalogue.  The stellar sources are those
with $\Delta_{2''-4''} \sim 0.1$--0.2.
Galaxies typically show larger values of $\Delta_{2''-4''}$.  
Our CCD imaging
readily differentiates between stars and galaxies down to $B\sim 20.5$, 
beyond the magnitude limit of the GRS/APM catalogue.
For the analysis of star-galaxy separation we use those sources
brighter than $B=19.5$.  Those with $\Delta_{2''-4''}<0.2$ are defined
as stellar.}
  \label{fig:aper}
\end{figure}

Following Reid et al.\ (1996) we construct a plot of the magnitude
difference between 4.0$''$ and 2.0$''$ diameter apertures
($\Delta_{2''-4''}$) on the $B$-band exposures versus total $B$
magnitude (Fig.~\ref{fig:aper}).  In such a diagram, stars should
trace a horizontal locus due to their fixed profile shape (the
telescope PSF) and therefore a fixed proportion of light in the two apertures.
The horizontal sequence is seen in Figure~\ref{fig:aper} at
$\Delta_{2''-4''}\sim 0.1$.  Galaxies, which exhibit a range
of more extended profile shapes, will show a wider range of
$\Delta_{2''-4''}$ values.  Close to the completeness limit of the frames,
the galaxy and star loci overlap due to low signal-to-noise.  Examination
of Fig.~\ref{fig:aper} reveals that our CCD imaging can readily differentiate
stars from galaxies down to a magnitude of at least $B\sim 20.5$.

\begin{figure*}
\centerline{\psfig{file=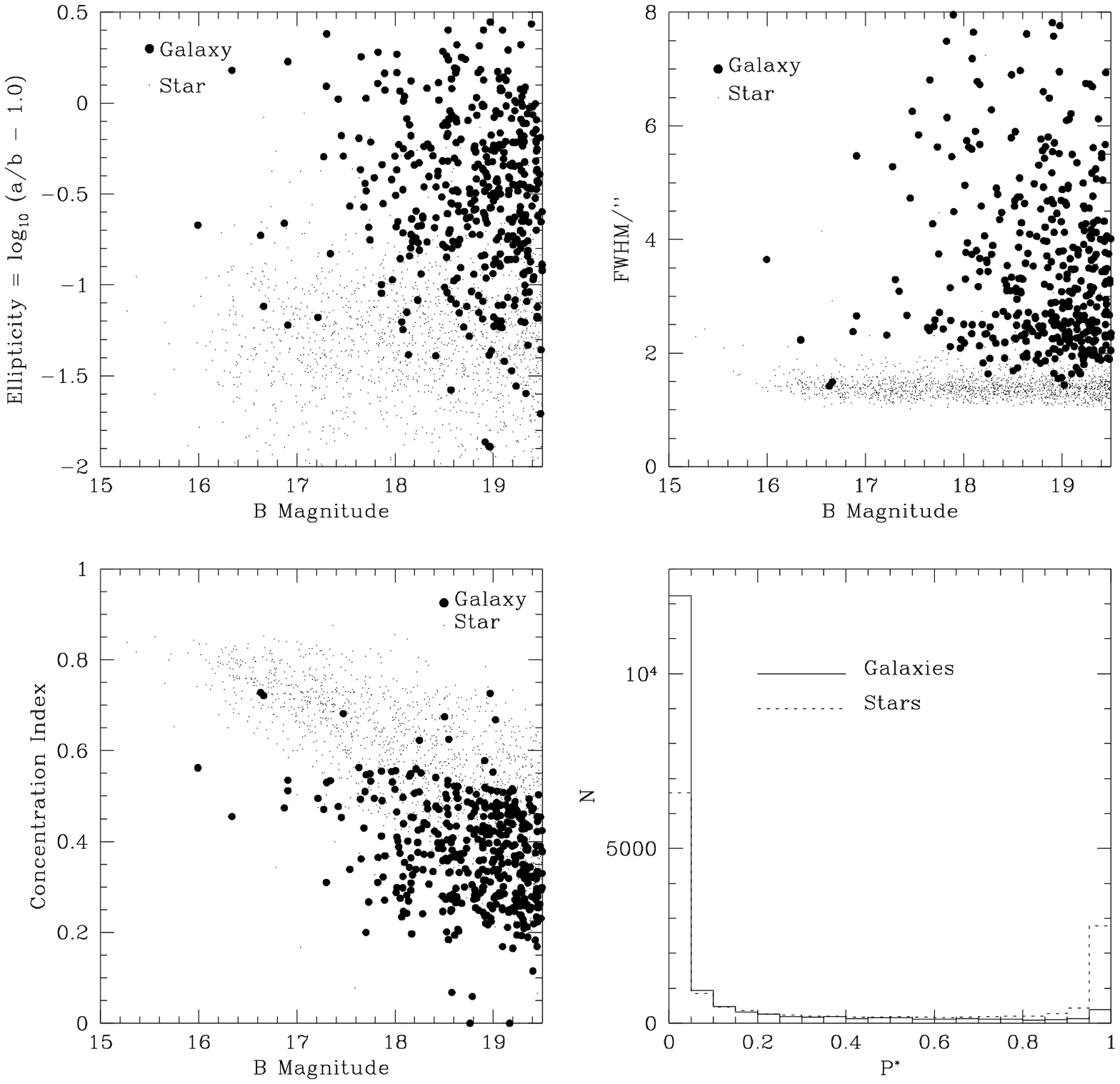,angle=0,width=5.0in}}
  \caption{Plots of critical SExtractor parameters for
stars and galaxies as defined from Figure~\ref{fig:aper}.  
Upper Left: Ellipticity
of the sources (defined as the ratio of semi-major to semi-minor axes)
as a function of total magnitude.  
Upper Right: FWHM as a function of total magnitude.  
Lower Left: Concentration index as a function of
total magnitude.  Lower Right: Histogram of $P^\ast$, SExtractor's own
probability of a given source being a star.}
\label{fig:stars}
\end{figure*}

We examine the potential of four further SExtractor parameters for usage
in star--galaxy separation.
Ellipticity is obtained from SExtractor's estimates of the semi-major
and semi-minor axes for sources.  The {\sc class\_star} parameter
($P^\ast$) is SExtractor's own estimate of stellarity and uses 
shape information fed into a neural-network classifier that has been
trained a priori on other images. 
Finally, we use two estimates of the compactness of the image profile:  
the concentration index (Abrahams et al.\ 1993) and FWHM.  
Plotted in Figure~\ref{fig:stars} are
these critical parameters for both stars and galaxies from Abell~22.
The other clusters follow the same distribution of sources on these
planes.  For the ellipticity and concentration index, the stars and
galaxies, as defined with the $\Delta_{2''-4''}$ cut, overlap somewhat,
whilst the converse is true for $P^\ast$ and the FWHM measurements.  
Therefore both the FWHM and $P^\ast$ are useful in differentiating
stars and galaxies.

Three criteria are used to generate a robust and 
stringent definition of galaxies: $\Delta_{2''-4''}>0.2$, 
FWHM $>2$''  and $P^\ast<0.1$.
In adopting these criteria, we must accept that a small number of
stars and galaxies will be mis-classified.  
However, as demonstrated in Figures~\ref{fig:aper} and \ref{fig:stars},
the number of stellar sources fulfilling these rigorous criteria is
small, $\leq 3$ percent. 

%
%
%

%
%
%

\section{Comparison of the GRS/APM Galaxy Catalogue with LARCS}

To test the reliability of the LARCS cataloging, photometry
and star--galaxy separation, we now compare our sample
to the galaxy catalogue from the APM survey, which is
used by the GRS/APM.

The APM survey (Maddox et al.\  1990a, 1990b) is constructed from
photographic plate materials taken at the UK Schmidt Telescope Unit
in Australia.  Approximately 390 photographic IIIa-J plates
are scanned by the APM.  The APM scans uses a
16-$\mu$m pixel with a sampling rate that produces a
0.54$''$ pixel scale.  Magnitudes are defined using the Kodak emulsion
in combination with a GG395 filter and zeropointed with Johnson $B$-band
CCD photometry.  The rms random error in each galaxy magnitude 
is 0.1--0.2 mags to a depth of $b_J = 20.5$ (Maddox et
al.\ 1990a).  Subsequent field corrections are applied to the APM
catalogues to account for non-photometric observations, vignetting and
differential desensitization.  Maddox et al.\  (1990b) argue that the
rms plate zero-point error in the final matched survey catalogue is
0.04\,mags.  Objects in the overlapping regions between the plates are
utilized for establishing the plate-to-plate variation and the 
zeropoints are iteratively corrected to remove these offsets.  

Image surface brightness profiles are used in the APM to achieve
star-galaxy separation.  The residual differences between a given image
profile and a stellar profile at the same magnitude are calculated, 
resulting in brighter stars being readily identified.  At 
fainter magnitudes, the lower signal-to-noise in the images makes
robust separation impossible.
Maddox et al.\ (1990a) estimate that the
catalogue's completeness is $\sim 98$ percent for $b_J<19.5$, with a stellar
contamination rate of 5--10 percent at the same depth depending upon
proximity to the galactic plane.

The 2dFGRS is an ambitious project to collect high quality redshifts and
spectra for $\sim$\,250,000 galaxies brighter than $b_J=19.45$
(extinction corrected).
Galaxies are robustly selected from the extended version of the APM galaxy
catalogue (Maddox et al.\ 1990a) covering a region of over 1700 square
degrees in both the Northern and Southern Pole regions and used
as the input to the 2dFGRS (Colless et al.\ 1998;
Maddox et al.\ 1998; Folkes et al.\ 1999).

Of the LARCS clusters, four overlap with the 2dFGRS parent catalogue from
the APM.  The details of these cluster catalogues
(Abell 22, 1084, 1650 and 1651)
are presented in Table~\ref{tab:global}.  The 0.3$''$ accuracy within
the LARCS astrometry is more than adequate for identifying source
matches with the APM catalogues.

%
%
\begin{table}
\begin{center}
\caption{Parameters of the LARCS clusters that
overlap with the GRS/APM, including the number of sources 
catalogued in the LARCS CCD frames, the number of galaxies 
(defined using the criteria presented in \S2.3) and 
the galactic cap region of the GRS/APM with which they overlap
(Southern Galactic Pole, SGP; Northern Galactic Pole, NGP). \hfil}
\begin{tabular}{lccl}
\noalign{\medskip}
\hline
Cluster & N(Objects) & N(Galaxies) & Overlap\\
\noalign{\medskip}\hline
Abell 22   & 31851 & 15429 & SGP \\
Abell 1084 & 37801 & 14494 & NGP \\
Abell 1650 & 34403 & 13480 & NGP \\
Abell 1651 & 42286 & 15801 & NGP \\
\hline
\noalign{\smallskip}
\end{tabular}
  \label{tab:global}
\end{center}
\end{table}

\subsection{Matching LARCS to GRS/APM}

Here, we compare the GRS/APM catalogue down to the 2dFGRS 
magnitude cut-off of
$b_J = 19.45$ (extinction corrected) across the LARCS/GRS overlap 
area of 12.3 square degrees.  

The APM catalogue is divided into two distinct areas: Northern and
Southern Galactic Pole regions.  There are 2754 GRS/APM
galaxies in the overlap of the GRS/APM with the four clusters
in Table~\ref{tab:global}.  One of the clusters (Abell~22) lies in the
Southern Galactic Pole Region, whilst the other three overlapping
clusters are in the Northern Galactic Pole Region.

%
%
\begin{table*}
\begin{center}
\caption{Comparison between the input GRS/APM catalogue
and the four full LARCS catalogues.  N(Matched) is the number
of GRS/APM sources that are matched to LARCS.
N(Galaxies) is the number of sources that we define as
galaxies using the criteria presented in \S3.  
N(Close blends) is the number of
galaxies that the LARCS survey detects that are blended 
or in close proximity to a secondary source, which 
affects the astrometric matching routine.  
N(APM noise) is the number
of sources existing within the GRS/APM survey that have no
likely counterparts
within LARCS -- they are random detections, perhaps a moving
source, plate flaw or a diffraction spike of a star.
\hfil}
  \label{tab:mats}
\begin{tabular}{lcccccc}
\noalign{\medskip}	\hline
Cluster & N(GRS/APM) & N(Matched) & N(Galaxies) & N(Close  & N(APM \\
        &        &            &             & blends)  & noise) \\
\noalign{\medskip}	\hline
Abell 22   & 634 & 611 & 553  & 23 & 0 \\
Abell 1084 & 473 & 463 & 371  & 10 & 0 \\
Abell 1650 & 681 & 630 & 518  & 48 & 3 \\
Abell 1651 & 966 & 932 & 752  & 34 & 0 \\
\noalign{\smallskip}	\hline
\end{tabular}
\end{center}
\end{table*}

The GRS/APM galaxy catalogue is matched to the full LARCS catalogues
using a search radius of 3$''$.  Virtually all
sources ($\sim 95$ percent) in the GRS/APM have corresponding matches within
LARCS.  Those sources for which matches are not found are all
visually inspected--their APM positions are over-plotted upon the
corresponding LARCS mosaic pointings.  In most cases
the unmatched sources result from differences in
blending and deblending of close sources between the two surveys.
Most are resolved in the LARCS CCD imaging, whereas the APM
sources appear as a blend of a number (typically 2--3) of LARCS
sources (both galaxy-galaxy and galaxy-star blends).  
The centroiding for the astrometric solution is thus
affected, resulting in the source lying outside the 3$''$ search radius.  
Only three sources are found within the GRS/APM catalogue 
that do not have any likely counterpart in the LARCS imaging.  
These very rare sources ($<0.1$ percent) are examples of APM
mis-detections perhaps caused by plate flaws
or moving/variable sources.  Table \ref{tab:mats}
summarizes the results of the matches with the total number of sources
within each catalogue.

We further investigate the nature of the GRS/APM sources that are
matched using the star--galaxy separation criteria of LARCS as
presented in \S2.  The number of matched sources which satisfy all
of the `LARCS galaxy criteria' are presented in Table
\ref{tab:mats}.  Broadly $80$ percent of matched sources defined as
galaxies by the GRS/APM are also classified as galaxies by LARCS.  
We note that $80  $ percent is the {\it lower} limit of the number of galaxies 
as our galaxy determination is very strict.

We are also able to place a lower limit on the 
stellar contamination by inverting the star-galaxy selection
criteria discussed previously.  Conservatively, stars 
can be defined as having $\Delta_{2''-4''}<0.2$, FWHM $<2.0$'' and 
$P^\ast>0.9$.  
Of the matched GRS/APM galaxies, $3  $ percent are defined as stars
by LARCS.  
Therefore at minimum, the stellar contamination of the APM is 3 percent,
whilst at maximum it is 20 percent.  
Our result is close to the original estimation of 5--10 percent stellar 
contamination by Maddox et al.\ (1990a).

\subsection{Photometric comparison}

%
%
\begin{figure}
\centerline{\psfig{file=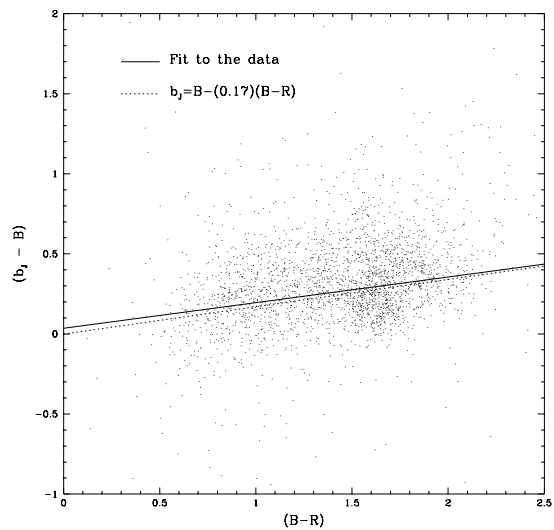,angle=0,width=3.0in}}
  \caption{
Plot of the correlation between ($b_J$-B) and (B-R) colour for 
sources from LARCS that are matched to the APM.  The solid 
line is the best fit to the data points.  The dotted line is the derived 
theoretical line $b_J=B-(0.17)(B-R)$.  The theoretical line is 
consistent with the fitted line.
}
  \label{fig:bjcorrel}
\end{figure}

In this section we examine the magnitudes of
sources in the LARCS and GRS/APM catalogues--searching for
zero-point differences between the four cluster fields, as well as
between the Northern and Southern Galactic Pole regions of
the GRS/APM.

To undertake such a comparison, one must convert LARCS' $R$-selected
sample to the $b_J$ system used by the APM.  To obtain $b_J$ magnitudes
we first convert the $R$-band magnitudes from SExtractor's
{\sc best\_mag} to total magnitudes.
The correction factor, 
$R_{tot} = {\sc best\_mag} - 0.021 \pm 0.010$,
is calculated from the median magnitude difference between SExtrator's 
{\sc best\_mag} magnitude and a large (15$''$ diameter) aperture
magnitude for isolated sources.  The correction is applied
to all sources to convert their {\sc best\_mag}
magnitudes to true total magnitudes.
Next, we convert these total $R$-band magnitudes, $R_{tot}$, to
total $B$-band magnitudes ($B_{\rm tot}$) using
$B_{\rm tot} = R_{\rm tot} + (B-R)$, 
where $(B-R)$ is the colour
measured from the seeing-matched tiles in a $4''$-diameter
aperture. 

The APM uses photographic $b_J$ magnitudes, and we
therefore require a colour transformation to convert our CCD-based
Johnson $B$-band magnitude into $b_J$.  Metcalfe et al.\ (1995) and a
number of other authors, including Colless (1989) and Bardelli et al.\
(1998), provide conversions from $B$ to $b_J$:  
$b_J = B - (0.28 \pm 0.04) (B-V)$.  
~From fitting the predicted colours of
galaxies across the range of redshift covered by the LARCS clusters
with non-evolving spectral energy distributions, we derive
the relation between
$(B-V)$ and $(B-R)$ to be $(B-R) = (1.66 \pm 0.03) (B-V)$.
Hence to convert from the $B$ magnitudes used in LARCS to the $b_J$ APM
magnitudes, we use $b_J = B - (0.17 \pm 0.05) (B-R)$.  
This conversion is consistent with the correlation between
$B-b_J$ and $(B-R)$ colour found for matched sources
between the APM and LARCS data (see Figure~\ref{fig:bjcorrel}),
where the calculated fit is $b_J = B - (0.16 \pm 0.03) (B-R) + (0.04\pm0.02)$.

In our comparison we use corrected (i.e.\ including a correction for 
galactic extinction, Schlegel et al. (1998))
$b_J$ magnitudes from the APM and corrected total $B$-band magnitudes from
LARCS, converted to $b_J$.  
Differences in the magnitudes of matched sources are examined as
functions of LARCS mosaic pointing and magnitude.

%
%
\begin{figure}
\centerline{\psfig{file=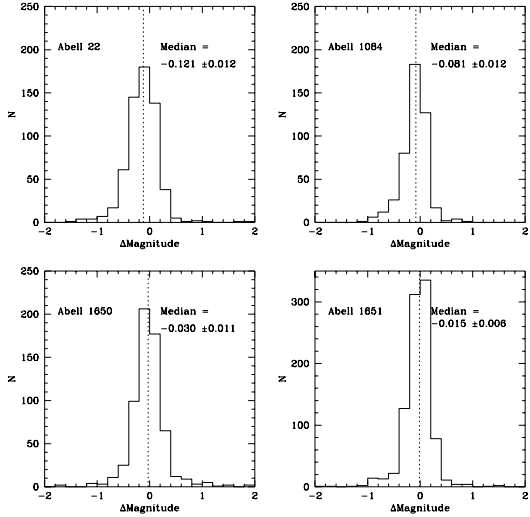,angle=0,width=3.0in}}
  \caption{
Histograms of the difference between APM magnitude and LARCS total
$b_J$ magnitude (the latter converted from $B$-band) for all four
clusters using the full LARCS catalogues.  
Abell~22 lies in the SGP, the other three
are in the NGP.  The median magnitude offsets and associated
$1\sigma$ errors are noted on the individual panels.}
  \label{fig:dmag1}
\end{figure}

Figure~\ref{fig:dmag1} shows the difference between LARCS $b_J$
and the total APM $b_J$ 
magnitude for each of the four clusters.  
The medians and the associated errors are also
indicated on the panels.  
Based on repeat scans of the plate material, Maddox et al. (1990a) 
suggest an intrinsic uncertainty in the zeropoint of individual
plates of 0.04 mag.
Of the four clusters, Abell~1650 and
Abell~1651 show very good agreement between the two photometric systems,
Abell~1084 shows a larger offset (but only at the $\sim 2.5\sigma$
level), while Abell~22 has a $\sim 3\sigma$ difference
($\Delta=0.12$\,mags).  We note that Abell~22 is the only one of the
four clusters which falls in the SGP area of the GRS/APM.

\begin{figure}
\centerline{\psfig{file=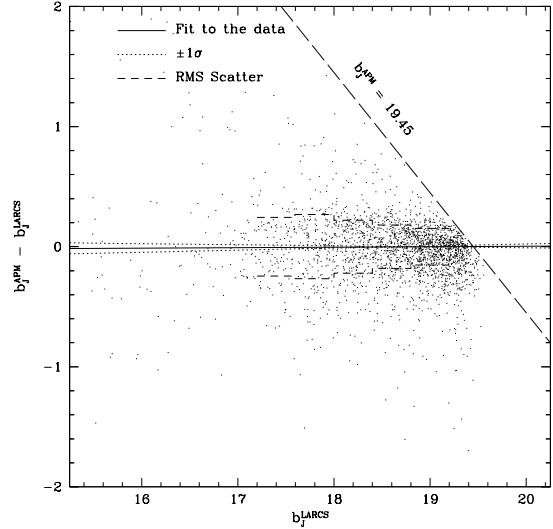,angle=0,width=3.0in}}
  \caption{
Difference in magnitude between LARCS and the APM as a function of
LARCS magnitude.
The solid line is a fit to those data points brighter than the median
magnitude ($b_J^{LARCS}=18.4$) so as to avoid the selection limit of 
the GRS/APM (long dashed line).  
The dotted line represents the $\pm 1\sigma$ limit of this fit.  
The RMS scatter (long dashed line) ranges from 0.18--0.25,
confirming the original estimate of 0.2 made by Maddox et al (1990a). 
}
  \label{fig:linear}
\end{figure}

We next investigate the precision of individual galaxy measurements.  
Maddox et al. (1990a) and Folkes et al. (1999) state that the random error
associated with the $b_J$ magnitudes is $\pm 0.2$ mag for the 
range $17.0 \leq b_J \leq 19.45$.
In Figure~\ref{fig:linear} we plot the magnitude difference in 
$b_J$ between
LARCS and the APM galaxies against LARCS magnitude.  Excluding regions of the plot
close to the selection limits, we find that the fit is consistent with a slope 
of zero, hence the APM demonstrates a linear photometric scale over at 
least the range $17 < b_J < 19$.
The RMS scatter is also consistent with the random error previously 
stated by Maddox et al (1990a).

%
%

In summary, therefore, we see slight evidence for systematic 
variations in the zero points of the GRS/APM and LARCS photometry 
across the four clusters, but such variations are within the limit 
imposed by the combined errors on the GRS/APM and LARCS zero points 
and thus are not significant.  We also find no 
evidence for systematic variations between the photometry of the 
SGP and NGP regions of the GRS/APM at levels $\gg 0.1$\,mag.  
We further note that there is no evidence for a magnitude offset 
between individual pointings of the mosaic tiles within LARCS at the
level of $> 0.02$\,mags (consistent with the photometric errors
on the sources used in the comparison).

\section{Comparison of the LARCS Galaxy Catalogue with the GRS/APM}

%
%
\begin{figure}
\centerline{\psfig{file=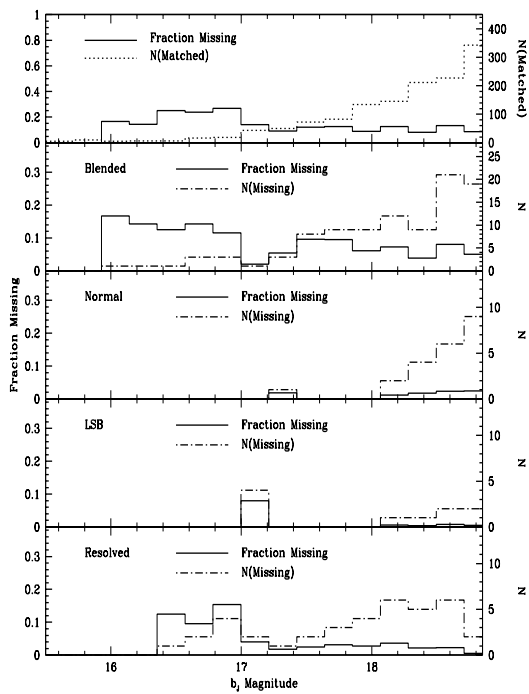,angle=0,width=3.0in,height=4.0in}}
  \caption{\small{
The fraction of bright LARCS galaxies missing from the 
GRS/APM as a function of $b_J$ magnitude.
The top panel shows the total fraction that is missing with
the total number of matched galaxies.  About 10--20 percent of
all galaxies are missed at all magnitudes.  The four lower panels
break down the missing population into the four broad constituent
types.  The normal and LSB populations are primarily
missed at $b_J > 18.0$, but the blended population
is missed across the magnitude range.  
See \S4.2 for discussion.
}}
  \label{fig:percents}
\end{figure}

%
%
\begin{table*}
\begin{center}
\caption{Comparison of the bright LARCS catalogues
with the GRS/APM catalogue.  N(Bright) is the number of sources within
the LARCS catalogues when cut at $b_J \ls 18.85$.
N(Matched) is the number of these LARCS sources matched to the
GRS/APM catalogue.  N(Galaxies, Matched) is the number of matched 
galaxies that fall within our stringent star-galaxy separation criteria,
whilst N(Galaxies, Unmatched) is the number of unmatched galaxies.
See \S4 for a full discussion.
\hfil}
\begin{tabular}{lcccccc}
\noalign{\medskip}	\hline
Cluster & N(Bright) & N(Matched) & N(Galaxies, & N(Galaxies, \\
 & & & Matched) & Unmatched) \\
\noalign{\medskip}	\hline
Abell 22   & 1997 & 249 & 226 & 35\\
Abell 1084 & 2904 & 177 & 152 & 29\\
Abell 1650 & 3016 & 328 & 249 & 35\\
Abell 1651 & 3521 & 471 & 381 & 70\\
\noalign{\smallskip}	\hline
\end{tabular}
  \label{tab:com2}
\end{center}
\end{table*}

We now compare the LARCS and GRS/APM catalogues in the reverse
sense: that is we construct a ``complete'' galaxy catalogue from 
LARCS and determine the overlap with the GRS/APM.  

Because the faint magnitude limit 
of the GRS/APM input catalogue is $b_J=19.45$, we
generate a bright LARCS sample for each of the four clusters.
As we confirmed in \S3.2, the typical photometric errors for galaxies
in the GRS/APM is 0.2 mag.  To ensure the most complete comparison
therefore we select all sources from the LARCS catalogues which
are at least $3\sigma$ (0.6 mag) away from the limit of the GRS/APM:
$b_J=19.45$.  This imposes a magnitude limit of $b_J=18.85$ and we list 
the number of sources in each field in this ``bright'' sample in
Table~\ref{tab:com2}.  
Our bright comparison is thus well above the completeness limit 
and the random scattering of sources above/below
the cut-off used for the GRS/APM catalogue.
Based upon the scatter at brighter magnitudes (see Figure~\ref{fig:linear})
selecting our sample at $b_J=18.85$ should exclude at most $\sim3$ percent 
of the population.
We further apply our conservative galaxy selection criteria (see \S2)
to the catalogues to create bright galaxy catalogues for comparison
to the GRS/APM input catalogue.

In principle, {\it all} of the galaxies contained within these
bright LARCS galaxy catalogues should be present in the GRS/APM.
Again, we employ a search radius of 3$''$ for the
matching algorithm and we give a summary of the results in 
Table~\ref{tab:com2}.

The GRS/APM identifies the majority of 
the galaxies contained within the bright LARCS galaxy sample.  
However, the GRS/APM is missing 10--20 percent
of the galaxies at all magnitudes:
Figure~\ref{fig:percents} presents a distribution
of the percentage of bright galaxies missing from the
GRS/APM as a function of $b_J$ magnitude, overlaid with
the total number of matched galaxies.
The missing population represents approximately a fixed 
fraction of the total population across the whole magnitude range.

To investigate the properties of the missing galaxy population we compare
the characteristics of the matched and unmatched
galaxies in terms of their morphologies, brightnesses, spatial distributions and $(B-R)$ colours.

\subsection{Morphologies of the Missing Populations}

\begin{figure*}
\centerline{\psfig{file=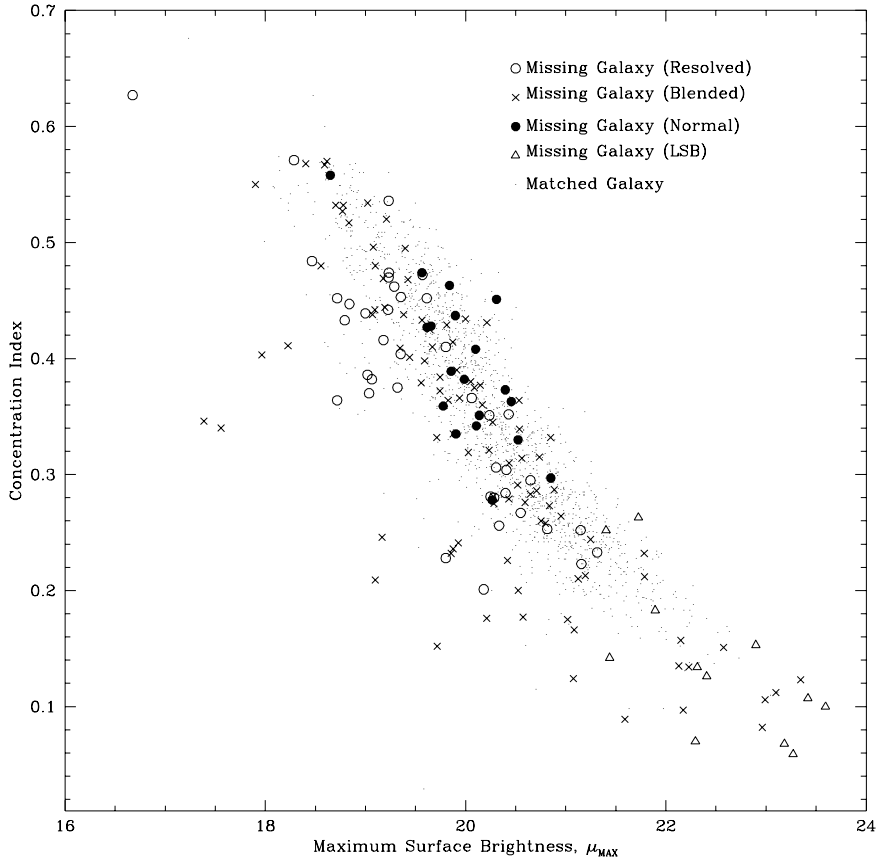,angle=0,width=5.0in}}
  \caption{\small{Peak surface brightness, $\mu_{MAX}$,
versus concentration index (from SExtractor) for the missing
galaxies and the matched sources (dots).
The missing population broadly traces the matched population save for the
LSBs, which preferentially lie to the lower-right. 
}}
  \label{fig:mu_ci}
\end{figure*}

Figure~\ref{fig:mu_ci} shows the distribution of maximum surface brightness
($\mu_{MAX}$) versus concentration index for the matched and
missing populations; late-types typically lie towards lower
right and early-types populate the upper left of this plane
(Abraham et al.\ 1994).
Figure~\ref{fig:mu_ci} demonstrates that, whilst the missing 
galaxy population broadly follows that of the matched population, 
there exist distinct classes of galaxies that are missed in 
the GRS/APM.

%
%

%
%
\begin{table*}
\begin{center}
\caption{Breakdown of the morphologies
of the bright ($b_J < 18.85$) `missing' galaxy population obtained
from visual inspection of the LARCS CCD material.  
N(Missing) is the total number of missing galaxies from the GRS/APM.
N(Blended) is the number of galaxies that are
blended or have a close companion and thus do not
appear as distinct sources in the APM.
N(Resolved) is the number of apparently normal, resolved galaxies 
missing from the GRS/APM.
N(Normal) is the number of small, often compact, normal galaxies 
(also see Drinkwater et al.\ 1999). 
These galaxies are mis-classified as stellar 
in the APM.  N(LSB) is the number of low surface brightness
galaxies missed by the GRS/APM.  
Several examples of each type of
missing galaxy are presented in Figure~\ref{fig:visual}.
\hfil}
\begin{tabular}{lcccccc}
\noalign{\medskip}	\hline
Cluster & N(Missing) & N(Blended) & N(Resolved) & N(Normal) & N(LSB)  \\
\noalign{\medskip}	\hline
Abell 22   & 35  & 22 & 5  & 6  & 2    \\
Abell 1084 & 29  & 17 & 10 & 1  & 1  \\
Abell 1650 & 35  & 18 & 8  & 4  & 4  \\
Abell 1651 & 70  & 42 & 16 & 7  & 5  \\
\noalign{\smallskip}	\hline
\end{tabular}
  \label{tab:com3}
\end{center}
\end{table*}

\begin{figure*}
\centerline{\psfig{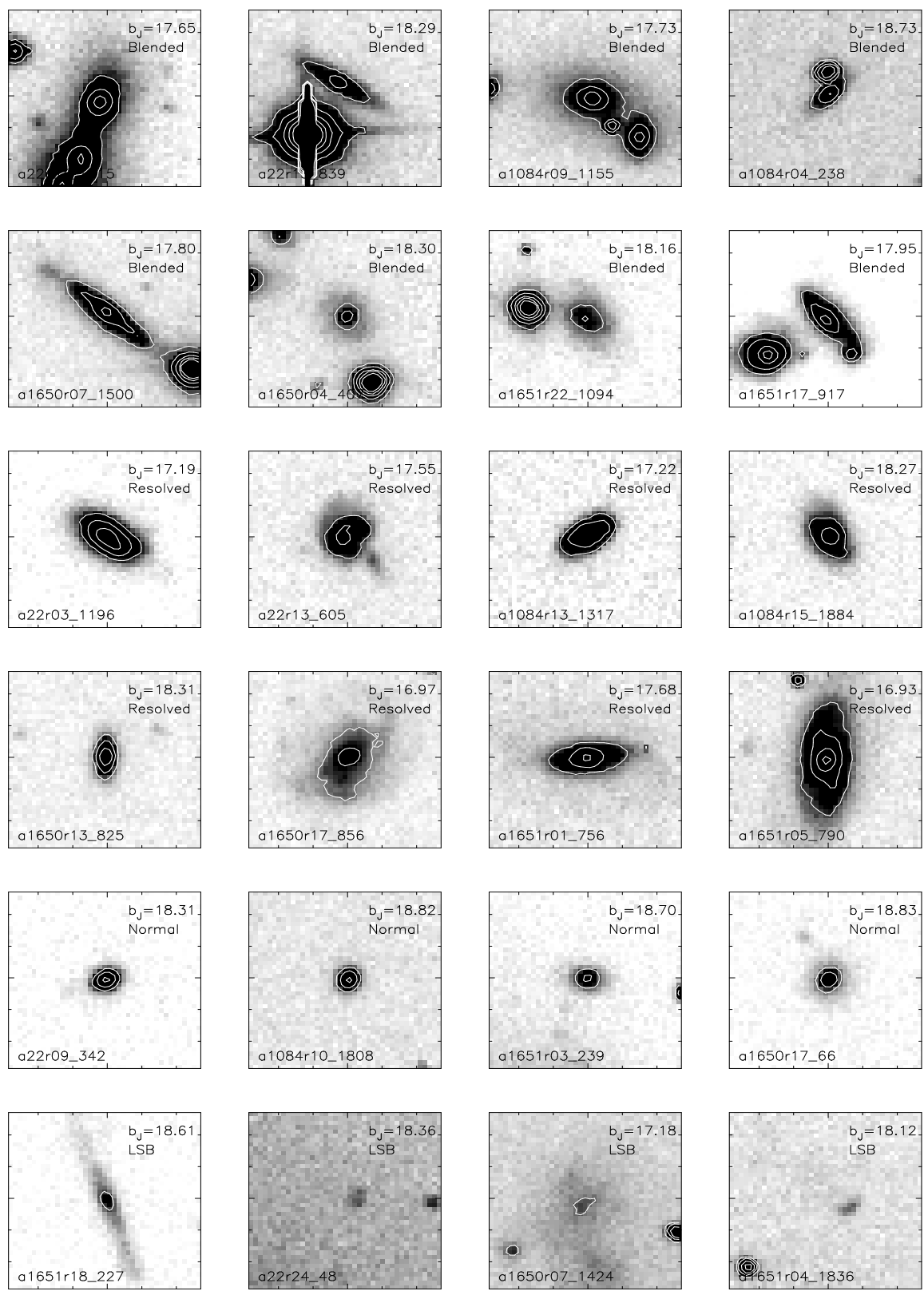}}
  \caption{\small{
Examples of galaxies missing from the GRS/APM catalogue that appear in 
the bright ($b_J < 18.85$) LARCS galaxy catalogues.  
(See \S4 and Table~\ref{tab:com3} for details of the bright comparison).
Each box is $30''$ on the side with tickmarks every $5''$. 
The contours start at a level of $\mu_B=22$\,mag.\ arcsec$^{-2}$ and
increase in brightness by 1\,mag arcsec$^{-2}$. 
The upper two rows are examples of blended galaxies near stars
and other galaxies.  The next two rows are normal looking, resolved
galaxies incorrectly classified in the APM as not `non-stellar'
in one or both APM passbands.  
The penultimate row are normal galaxies, classified as stars in the
APM.  The last row are low surface brightness galaxies.  
}}
  \label{fig:visual}
\end{figure*}

Figure~\ref{fig:visual} displays a sample of some of these bright 
galaxies missed from the GRS/APM.
We visually classify the unmatched galaxies and define four broad 
classes absent from the GRS/APM:

\begin{itemize}
\item Blends.  There are many examples of blended galaxies (see the 
top two rows of Figure~\ref{fig:visual}).
Due to the proximity of a close neighbour or neighbours 
(either other galaxies or stars), 
the galaxy is excluded from the GRS/APM catalogue.

\item Resolved.  There are several apparently `normal', resolved
galaxies missing from the GRS/APM (see the third and fourth rows of
Figure~\ref{fig:visual}).

\item Normal.  There are also small, mostly compact, galaxies
which have been missed, the bulk of these are likely to
be mis-classified as stellar in the APM and missed from the galaxy
survey input catalogue (see the penultimate row of Figure~\ref{fig:visual}).  

\item LSB.  There are a small number of low
surface brightness galaxies (see the bottom row of Figure~\ref{fig:visual}).
\end{itemize}

%
%

Blends are the dominant component of the missing galaxy population, 
comprising $60  $ percent of it, resolved galaxies $20  $ percent, normal galaxies $15  $ percent and low surface
brightness galaxies $5  $ percent (see Table~\ref{tab:com3}).

The majority of the missing population are galaxies blended with a
secondary source; both with galaxies (e.g.\ a1084r09\_1155 in Figure
~\ref{fig:visual}) and stars (e.g.\ a22r13\_839).
In one or both of the APM passbands (http://www.ast.cam.ac.uk/$\sim$apmcat/)
these blends have a {\sc merged} flag set.
Moreover, as the APM data are scanned from many different plates, the plate
quality varies from cluster to cluster.
In principle, the plates should be of top quality, but subjective
grading of plate quality and variable focus would cause 
the median separation between blended sources to vary and hence 
affect the success of deblending.  
Table \ref{tab:blenddist} presents
the median separation of the missing, merged galaxies.
Clearly there are differences in plate quality (and hence
the deblending) across 
the APM, with one plate (containing Abell 1084) having a 
significantly smaller median blend offset than other plates 
and the fewest number of blends.

There exists a number of apparently `normal', resolved galaxies that 
the GRS/APM misses (e.g.\ a22r03\_1196). 
These account for $\sim 20$ percent
of the missing populations.  We investigate these resolved
galaxies by inspecting them in the online APM catalogues
(see http://www.ast.cam.ac.uk/$\sim$apmcat).  The Resolved
galaxies have classification flags set in
one or both passbands to `noise', `merged' 
or `stellar' (e.g.\ a1651r05\_790 has
classification flags set to `non-stellar' and `stellar' 
in the B and R-scans of the APM). 

Whilst issues of blending account for $\sim60$ percent of the missing APM
galaxies, $\sim15$ percent of the remainder are bright and relatively
compact in nature.  With the poorer resolution of the APM, these
galaxies are likely to be classified as stellar and indeed most are
typed as `stellar' according to their classification flag in the APM
catalogues (e.g.\ a1084r03\_50 has
classification flags in both passbands set to `stellar').
Drinkwater et al.\  (1999) found evidence for a similar population
of small, high surface brightness, compact galaxies within the Fornax
cluster.  They argue that such galaxies are readily overlooked in many
surveys because they are simply misclassified as stars.  We confirm
that the GRS/APM is systematically missing these 
compact galaxies from the galaxy catalogue.

The remaining $\sim 5$ percent of the missing galaxies are 
low surface brightness
systems (LSB).  The few examples of
the LSB class that are found at all in the on-line APM catalogues
are classified as `noisy', but the majority are missing
(e.g.\ a1651r18\_227 has classification
flags set to both `noise-like' and `merged' in the red and blue
APM passbands respectively).
Cross et al.\ (2000) present the bivariate brightness distribution 
(BBD) for the 2dFGRS survey, which confirms that the majority of these 
LSB galaxies found in LARCS are outside 2dFGRS's detection limits 
in surface brightness.

%
%
\begin{table}
\begin{center}
\caption{Median distance between the two closest neighbours in
a merged source with bootstrap estimates of the $1 \sigma$ errors.  
These results may be indicative of varying 
success of deblending in the APM.
\hfil}
\begin{tabular}{lccl}
\noalign{\medskip}
\hline
Cluster & Median Merger \\
        & Distance ($''$) \\
\noalign{\smallskip}\hline
Abell 22   & $8.59 \pm 0.85$ \\
Abell 1084 & $5.34 \pm 0.88$ \\
Abell 1650 & $7.28 \pm 0.65$ \\
Abell 1651 & $7.49 \pm 0.53$ \\
\hline
\noalign{\smallskip}
\end{tabular}
  \label{tab:blenddist}
\end{center}
\end{table}

\subsection{Magnitude Distribution of the Missing Populations}

The missing galaxies appear to be a fixed fraction, about 10--20 percent, 
of the total population at all magnitudes (see Figure~\ref{fig:percents}, 
upper panel).  The proportion of missing galaxies at the brightest 
magnitudes, $b_J<17$, increases to 20 per cent.

Because the missing Blended population dominates the total 
missing population number, these objects exhibit the same behaviour as the whole 
missing population across the entire magnitude range.
There is no strong dependence on brightness when galaxies are 
lost from the GRS/APM due to blending.

The missing Resolved galaxies occur more frequently
at brighter magnitudes, with few if any fainter than $b_J\sim 18.7$.
Many of these brighter normal galaxies are misclassified
in one or both APM passbands as blends due to slight 
asymmetries or irregularities in their morphology.

In contrast, the missing Normal and LSB
galaxies are primarily found at fainter magnitudes, $b_J >18$,
relatively uniformly distributed between $b_J\sim 18$ and the magnitude 
limit of the GRS/APM.  Each of these two sub-classes makes up 
about 1--2 percent of 
the total population at $b_J >18$ or fainter. 
The sensitivity limits of the plate material used in the
APM survey means that it preferentially misses faint LSBs,
but the dominant factor for distinguishing between 
faint compact galaxies and stars is the spatial resolution of
the APM.

\subsection{Spatial Distribution of Missing Populations}

%
%
\begin{figure}
\centerline{\psfig{file=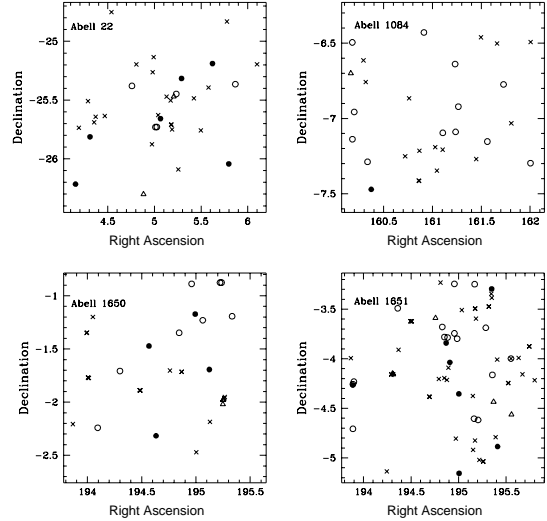,angle=0,width=3.0in,height=3.0in}}
  \caption{\small{
The spatial distribution of the galaxy populations
missing from the GRS/APM.
Each sub-population, Blends (crosses), Compact (solid circles),
Normal (open circles) and LSBs (open triangles) is 
scattered throughout the two degree fields, with only a slight 
over-density towards the centre of the cluster (located in 
the centre of each plot).  As Figure~\ref{fig:radial} shows, the over-density simply 
reflects the higher proportion of galaxies present in the 
cluster core.
}}
  \label{fig:distrib}
\end{figure}

\begin{figure}
\centerline{\psfig{file=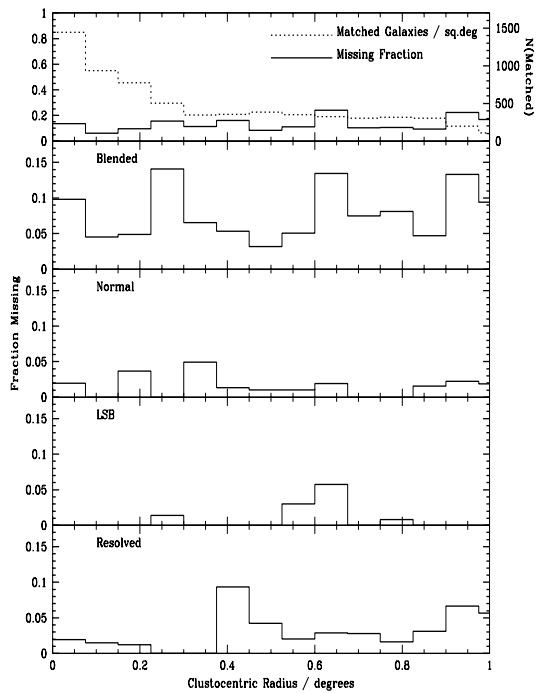,angle=0,width=3.0in,height=4.0in}}
  \caption{\small{
Number of galaxies matched ($b_J \leq 18.85$) per square degree as a function of
clustocentric radius, over-plotted with the total fraction of missing
galaxies (top panel).
The missing populations are approximately a fixed fraction
of the total population, which itself decreases with increasing
clustocentric distance. 
The lower four panels illustrate the contribution to the missing
fraction for each of the four sub-classes.
}}
  \label{fig:radial}
\end{figure}

The spatial distribution of the missing galaxies is presented
in Figure~\ref{fig:distrib}.
The galaxies are spread evenly throughout the 
four LARCS clusters.
We show the radially-averaged distribution of the
different missing galaxy classes for the combined sample from the
four clusters (Figure~\ref{fig:radial}).  We see that the 
missing galaxies represent a constant fraction 
of the whole population at all radii.  
Although it might be expected that in the crowded  
clusters cores the proportion of blended galaxies missing from
the GRS/APM would increase there is no evidence
for this.

%
%

\subsection{Colours of the Missing Populations}

\begin{figure}
\centerline{\psfig{file=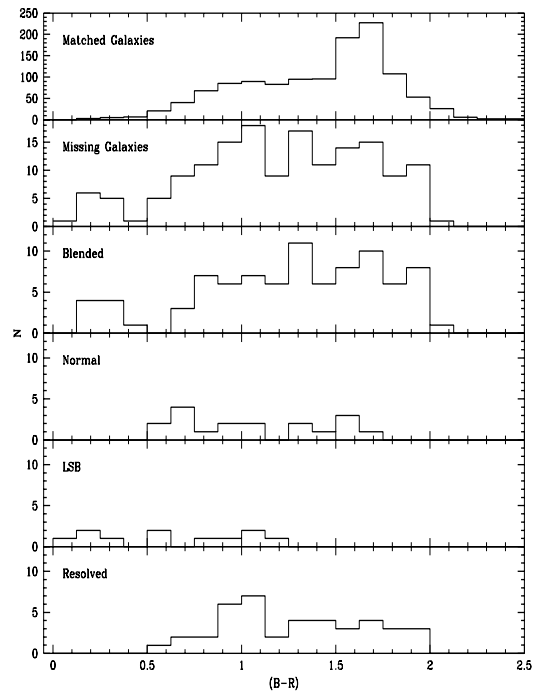,angle=0,width=3.0in,height=4.0in}}
  \caption{\small{
The $(B-R)$ colour histograms of the matched and
missing galaxy populations, with the missing population sub-divided
into its various morphological classes.  
Note the excess of matched galaxies with $(B-R)=1.6$ -- early-type 
cluster members that account for only around 10 percent of the galaxies 
in the comparison.
}}
  \label{fig:colorhists}
\end{figure}

Histograms of the $(B-R)$ colours of the `missing' population are 
presented in Figure~\ref{fig:colorhists} together with the colours 
of the `matched' population.  
The excess $(B-R)=1.6$ galaxies in the matched population are
the early-type cluster members; they account for only $\sim10$ percent of
the galaxies matched in the bright comparison. 
The missing galaxies span the whole
range in colours seen in the matched population, with 
a slight bias against galaxies with colours similar to early-type
galaxies in the clusters and a slightly enhanced number of blue galaxies.
The population of Blends dominates the total missing
galaxy population.  It broadly follows the matched population except
for a slight (but not significant) enhancement in the population
of very blue galaxies.
The Resolved population of missing galaxies traces that of the 
matched fairly well.
Finally, the distributions of missing Normal and LSB galaxies both
show shifts to the blue compared to the matched population.
This is more pronounced for the missing LSBs, but is still statistically
significant for the missing Normal population (likelihood $< 10^{-4}$).

%
%

We conclude that the colour biases in some of the populations
missing from the GRS/APM are relatively modest as compared
to the matched galaxies, especially for
the Blended population which dominates the missing sample.

\section{Conclusions}

We present the imaging dataset for the Las Campanas and
Anglo-Australian Observatories Rich Cluster Survey. This
programme is a panoramic CCD-based imaging and spectroscopic study of
21 of the most X-ray luminous clusters of galaxies at
$z=0.07$--0.16.  Our imaging extends out to radii of 10\,Mpc at
the median redshift of our survey--sufficient 
to encompass the infall region around
the clusters.  We describe the details of the
data reduction pipeline and calibration; our photometry is accurate to
$\leq 0.03$\,mags and our astrometry to better than 0.3$''$, as determined
from cross-comparisons between overlapping mosaic tiles.  We develop a
reliable star-galaxy separation technique to provide a low ($\sim2$ percent)
stellar contamination in the final galaxy catalogues.

We compare the results from the LARCS survey for four fields
that overlap with the 2dFGRS survey, which is based upon photographic plates
scans from the APM.  The comparison covers an area $\sim 12.3$ square
degrees on the sky and includes roughly 2500 galaxies brighter than
$b_J=19.45$.  

The photometric accuracy of LARCS is accurate to better than 0.03 mag
with the internal tile-to-tile photometric match being $\leq 0.005$ mag.
There is no evidence for a significant offset 
($\gg 0.1$mag) between fields in the LARCS survey or between the 
disconnected regions of the Northern and Southern APM catalogues. 

The APM magnitudes are linear across at least two magnitudes
in $b_J$ and the random photometrical errors are found to be consistent
with the value of 0.2 mag estimated by Maddox et al (1990a).

The stellar contamination within the GRS/APM is found to be 
about 3--20 percent, close to the range of 5--10 percent
originally estimated by Maddox et al.\  (1990a).

The number of false detections within the APM is very small, at the
$\sim0.1$ percent level.

There are several populations of galaxies that are not
present within the APM input catalogue: these comprise roughly 10--20 percent
of the total galaxy population at all magnitudes.  
The missing galaxies can be divided in to four broad
categories: Blends ($60$ percent), Resolved galaxies ($20$ percent), 
Normal galaxies ($15$ percent) and Low Surface Brightness 
galaxies ($5$ percent).

\begin{itemize}
\item The Blends can readily be explained.  Due to the resolution of the
photographic plates, the APM has not been able to cleanly deblend two
sources.  As a result these sources are flagged as `Merged' and
removed from the GRS/APM input catalogue.  The higher spatial
resolution and dynamic range available in the CCD-based LARCS survey
allows us to reliably deblend these systems and to investigate the
properties of the missing galaxies.

\item The missing population of Resolved galaxies has 
classification flags set in
one or both of the APM passbands as
stellar, merged or noise.  Therefore they are not included in the input
catalogue of the 2dFGRS.

\item The relatively compact Normal galaxies missing from 
the GRS/APM input catalogue are classified as stars due to 
the modest spatial resolution of the plates.

\item The Low Surface Brightness galaxies are not present as the APM 
mostly classifies them as `noisy' if they are detected.  Indeed, the work
of Cross et al.\ (2000) suggests that the majority of them will lie beyond
the surface brightness detection limit of the 2dFGRS. 
These galaxies only represent 5 percent of a magnitude-limited sample 
and so their absence is less of a concern than the other classes 
of missing galaxies.

\end{itemize}

To summarise, the APM input catalogue to the 2dFGRS survey has
a stellar contamination of about 5--10 percent as originally estimated, but misses
10-20 percent of all galaxies at all magnitudes.  In terms of their
magnitudes, surface brightnesses, colours and compactness, the majority
(80 percent of the missing fraction, or $\sim 10$ percent of the total galaxy
population) of these missing galaxies are similar to those included
in the GRS/APM and so their absence could be corrected for by simply
assuming they represent a random selection of the matched population.
However, the remaining 5 percent of the  missed galaxy population 
consists of either LSB or Normal galaxies--these have
significantly different properties than the matched population and
so their absence cannot be easily corrected for in the 2dFGRS catalogue.

This paper is the first in a series based upon the LARCS survey.  Our
next paper will focus on the variation in the colours  of the cluster
population with environment (Pimbblet et al.,\ in prep).

\subsection*{Acknowledgments}

We thank Simon Driver and Nicholas Cross
for getting the APM catalogue to us.   
We thank Shaun Cole and Peder Norberg for useful 
and stimulating discussion.
KAP acknowledges support from his PPARC studentship.
IRS and ACE acknowledge support from Royal Society fellowships.
WJC and EOH acknowledge the financial support of the
Australian Research Council throughout the course of this work.
AIZ acknowledges support from NASA grant HF-01087.01-96.
We thank the the Observatories of the Carnegie Institution of
Washington for their generous support of this survey.
This research has made extensive use of the University of Durham's 
{\sc STARLINK} computing facilities and the facilities at Las
Campanas Observatory.

\clearpage



\medskip


\begin{thebibliography}{}


\bibitem[Abraham, Valdes, Yee \& van den Bergh(1994)]{1994ApJ...432...75A} 
Abraham R.G., Valdes F., Yee H.K.C., van den Bergh S., 1994, 
\apj, 432, 75 

\bibitem[Bardelli et al.(1998)]{1998MNRAS.296..599B} Bardelli S., Zucca 
E., Zamorani G., Vettolani G., Scaramella R., 1998, \mnras, 296, 599 

\bibitem[Bertin \& Arnouts(1996)]{1996A&AS..117..393B} Bertin E., 
Arnouts S., 1996, \aaps, 117, 393 

\bibitem[Cross et al.(2000)]{2000MNRAS} Cross N., Driver S.P., 
Couch W.J., et al, 2000 (submitted).

\bibitem[Colless(1989)]{1989MNRAS.237..799C} Colless M., 1989, \mnras, 
237, 799 

\bibitem[Colless(1998)]{1998wfsc.conf...77C} Colless M., 1998, Wide Field 
Surveys in Cosmology, 14th IAP meeting held May 26-30, 1998, Paris.\ 
Publisher: Editions Frontieres.\ ISBN: 2-8 6332-241-9, p.\ 77



\bibitem[Drinkwater et al.(1999)]{1999ApJ...511L..97D} Drinkwater M.J., 
Phillipps S., Gregg M.D., Parker Q.A., Smith R.M., Davies J.I., 
Jones J.B., Sadler E.M., 1999, \apjl, 511, L97 

\bibitem[Ebeling et al.(1996)]{1996MNRAS.281..799E} Ebeling H., Voges W., 
Bohringer H., Edge A.C., Huchra J.P., Briel U.G., 1996, 
\mnras, 281, 799 (XBACs)


\bibitem[Folkes et al.(1999)]{1999MNRAS.308..459F} Folkes S., et al., 
1999, \mnras, 308, 459.  (also see http://www.mso.anu.edu.au/2dFGRS/) 

\bibitem[Glazebrook, Peacock, Collins \& Miller(1994)]{1994MNRAS.266...65G} 
Glazebrook K., Peacock J.A., Collins, C.A., Miller L., 1994, 
\mnras, 266, 65 (G94)

\bibitem[Kibblewhite et al.(1984)]{1984astt.coll...89K} Kibblewhite E., 
Bridgeland M., Bunclark P., Cawson M., Irwin M., 1984, IAU Colloq.\ 
78: Astronomy with Schmidt-Type Telescopes, 89 

\bibitem[Landolt(1992)]{1992AJ....104..372L} Landolt A.U., 1992, \aj, 
104, 372 

\bibitem[MacGillivray \& Stobie(1984)]{1984VA.....27..433M} MacGillivray 
H.T.\ \& Stobie R.S., 1984, Vistas in Astronomy, 27, 433 

\bibitem[Maddox, Efstathiou, Sutherland \& Loveday(1990)]
{1990MNRAS.243..692M} 
Maddox S.J., Efstathiou G., 
Sutherland W.J., Loveday J., 1990a, \mnras, 243, 692.
(also see http://www.ast.cam.ac.uk/$\sim$apmcat/)

\bibitem[Maddox, Efstathiou \& Sutherland(1990)]{1990MNRAS.246..433M} 
Maddox S.J., Efstathiou G., Sutherland W.J., 1990b, \mnras, 246, 
433 

\bibitem[Maddox et al.(1998)]{1998lsst.conf...91M} Maddox S., et al., 
1998, Large Scale Structure:  Tracks and Traces, 91 

\bibitem[Metcalfe, Fong \& Shanks(1995)]{1995MNRAS.274..769M} Metcalfe N., 
Fong R., Shanks T., 1995, \mnras, 274, 769 


\bibitem[O'Hely et al.(1998)]{1998PASA...15..273O} O'Hely E., Couch W.J., 
Smail I., Edge A.C., Zabludoff A., 1998, PASP, 15, 273 

\bibitem{EOH PhD} O'Hely E., PhD Thesis `Bridging the gap', UNSW, 2000.

\bibitem[Reid et al.(1996)]{1996AJ....112.1472R} Reid I.N., Yan L., 
Majewski S., Thompson I., Smail I., 1996, \aj, 112, 1472 

\bibitem[Schlegel, Finkbeiner \& Davis(1998)]{1998ApJ...500..525S} 
Schlegel D.J., Finkbeiner D.P., Davis M., 1998, \apj, 500, 525 

\end{thebibliography}
\end{document}